\documentclass[]{aa}  

\usepackage{graphicx}
\usepackage{lineno}
\usepackage{epstopdf}
\usepackage{txfonts}
\usepackage{subcaption}         
\usepackage{lscape}             
\usepackage{placeins}           
\usepackage{hyperref}
\usepackage{amsmath}	
\usepackage{amssymb}	
\usepackage{upgreek}
\usepackage{xcolor}
\usepackage{textcomp}
\usepackage{natbib}

\newcommand{\orcid}[1]{\href{https://orcid.org/#1}{\includegraphics[width=10pt]{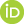}}}
\linenumbers
\renewcommand\makeLineNumber{}

\defcitealias{kangas22}{K22}
                                
\begin{document}

   \title{Blackbody evolution and circumstellar medium properties of broad-lined H-rich superluminous supernovae}

   \subtitle{}


   \author{T. Kangas
          \inst{1,2}\orcid{0000-0002-5477-0217}
          \and
          T. Nagao\inst{3}\orcid{0000-0002-3933-7861}
          \and
          U. Pylypenko\inst{1,2}\orcid{0009-0002-7560-1903}
          \and
          S. Schulze\inst{4}\orcid{0000-0001-6797-1889}
          \and
          L. Yan\inst{5}\orcid{0000-0003-1710-9339}
        }

   \institute{Finnish Centre for Astronomy with ESO (FINCA), FI-20014 University of Turku, Finland;
            \email{tjakangas@gmail.com}
            \and
            Department of Physics and Astronomy, FI-20014 University of Turku, Finland 
            \and
            National Astronomical Observatory of Japan, National Institutes of Natural Sciences, 2-21-1 Osawa, Mitaka, Tokyo 181-8588, Japan
            \and
            Department of Particle Physics and Astrophysics, Weizmann Institute of Science, 234 Herzl St, 76100 Rehovot, Israel
            \and
            The Caltech Optical Observatories, California Institute of Technology, Pasadena, CA 91125, USA
            }

   \date{Received October 12, 2026}

 
  \abstract
   {Broad-lined H-rich superluminous supernovae (SLSNe~II) tend to show signs of interaction, including an ultraviolet (UV) excess compared to a blackbody function in their spectral energy distribution (SED). However, a quantitative analysis of the CSM configuration and the UV excess has so far not been performed for most of the literature sample. Additionally, recently, one SLSN~II (SN~2023gpw) showed signs of asymmetry in its peculiar blackbody and spectral evolution, raising the question of whether similar properties can be found in other SLSNe~II or whether SN~2023gpw is an outlier in the sample.}
   {We aim to characterize the circumstellar environment of SLSNe II by searching for signs of asymmetry, quantifying the UV excess and examining its possible origins, and investigating any infrared (IR) echoes produced by circumstellar dust. Diversity within the literature sample of SLSNe~II is also explored.} 
   {We fit blackbody functions to the SEDs of SLSNe~II, then subtract them from UV data, when available, to quantify the UV excess luminosity and its evolution. We use this to investigate mass-loss rates required to produce the excess through interaction. We also investigate the line profiles in SLSN~II spectra at late times and compare line velocities to the blackbody radius evolution. We examine public \textit{WISE} IR images and place constraints on dust-related IR echoes.}
   {We find mass loss rates  of $\sim10^{-4}$ to $\sim0.03\, \mathrm{M_\odot}$~yr$^{-1}$ over the last years or decades of SLSN~II progenitor stars when assuming all interaction luminosity emerges in the UV, making these values conservative lower limits for the total mass loss rate. We find that radii calculated from line velocities in late-time spectra are mostly too large to come from a spherical interaction photosphere, and these spectra are only compatible with the blackbody radius evolution if the CSM is non-spherical with a low filling factor. This is also in some cases supported by a mismatch between spectroscopic and blackbody expansion velocities. However, considerable diversity is seen in the sample, probably as a result of different viewing angles and CSM masses. \textit{WISE} images seem too shallow to observe IR echoes in SLSNe~II beyond $z\sim0.1$, but in the nearest event, SN~2020jhm, a likely IR echo (which would be undetectable at higher redshifts) is detected, suggesting pre-existing dust.}
   {}


   \maketitle

\section{Introduction}
\label{sec:intro}

Massive stars ($\gtrsim8~\mathrm{M}_\odot$) end their lives in luminous explosions known as supernovae (SNe), divided into two main classes: hydrogen-poor SNe~I and hydrogen-rich SNe~II. For both types, there is also an analogous type of superluminous supernovae (SLSNe) -- for reviews on SLSNe in general, see \citet[][]{galyam12,galyam19,moriya24} -- which reach peak absolute magnitudes of $\lesssim-20$~mag. This limit is the rough border between normal stripped-envelope SNe and spectroscopic SLSNe~I \citep{decia18,galyam19}, while for H-rich SLSNe such a limit is more arbitrary. For recent sample studies, see, e.g., \citet{chen23a,chen23b} and \citet{pessi24}, for H-poor and H-rich SLSNe respectively. H-rich SLSNe with narrow emission lines, hereafter referred to as SLSNe~IIn, have been shown to be the extension of ``normal" slowly-evolving SNe~IIn to higher luminosities \citep{hiramatsu24}, and the distribution of properties between the two is continuous. Similarly, H-rich SLSNe whose spectra do not show strong narrow features, hereafter referred to as SLSNe~II, mostly show spectra similar to normal SNe~II.

SLSNe~II are rare compared to other SLSN types, and not many studies of them exist so far. The prototypical event of this class was SN~2008es \citep{gezari09,miller09}, which was argued to resemble a SN~II-L, though much more luminous and slowly evolving. Sample studies have been performed by \citet{inserra18} and \citet{kangas22}; the latter of these, hereafter also \citetalias{kangas22}, is the largest SLSN~II sample in the literature so far, with 13 events \citep[the 14th was later shown to be a tidal disruption event by][]{yao23}. In these studies, it was shown that despite the lack of narrow lines, SLSNe~II still exhibit signs of circumstellar interaction (CSI); these include multi-component H$\alpha$ profiles seen in some SLSNe~II \citep{inserra18,hebbar26} and ultraviolet (UV) excesses over blackbody functions in their spectral energy distributions (SEDs). It is possible that, similarly to SLSNe~IIn, at least some SLSNe~II are simply the most strongly interacting SNe~II that, possibly because of a non-spherical circumstellar medium (CSM), avoid the narrow spectral lines of SNe~IIn \citep[see, e.g.,][]{smith15,nagao25,reynolds25b}.

However, the light curves of SLSNe~II are mostly smooth, with no bumps that would indicate episodic mass loss, and compatible with models of a magnetar spin-down \citep{kb10} central engine \citep{inserra18,kangas22}. The energy emitted by some of the most luminous SLSNe~II, spectroscopically similar to the prototypical SN~2008es, can be on the order of $10^{51}$~erg or more, sometimes even exceeding the maximum explosion energy of a typical neutrino-driven SN \citep[$\sim2\times10^{51}$~erg;][]{janka12}, indicating that the conversion of this kinetic energy into emission through CSI, while present, is insufficient to explain all of them \citep{kangas22,kangas26}. The host galaxies of these extreme, 2008es-like SNe also show a similarity to those of magnetar-powered SLSNe~I \citepalias{kangas22}, supporting this connection but (due to a small sample size) not confirming it. In a few SLSNe~I, late-time broad H$\alpha$ emission from CSI \citep{yan15,yan17,pursiainen22} has also been seen, raising the question of whether there could even be a continuum from SLSNe~I to some magnetar-driven SLSNe~II through increasing interaction. Some SLSNe~I also exhibit resonance absorption features from CSM shells \citep{lunnan18,gkini26}, but this CSM is likely to be H-poor.

It is not clear whether some less-extreme SLSNe~II also have a central engine, but the 2008es-like SNe seem to form their own small subclass -- though a borderline case, SN~2023gpw \citep{kangas26}, complicates matters by having an emitted energy of $\sim10^{51}$~erg, but exhibiting a light-curve drop that the standard magnetar model has trouble explaining and not being spectroscopically 2008es-like. As for SN~2008es itself, \citet{kornpob19} showed that its late light curve was more consistent with being interaction-dominated but that a magnetar power source could not be ruled out.

The progenitor stars of SLSNe~II are, by necessity, massive and either H-rich or at least surrounded by H-rich CSM, but remain otherwise poorly understood. In this paper, we return to the SLSN~II sample of \citetalias{kangas22}, bolstered with SNe~2008es and 2023gpw, to examine aspects of the data not covered in detail in \citetalias{kangas22} and to shed light on the CSM properties in SLSNe~II such as asymmetry and mass-loss rates, which have implications on the late stages of these progenitors. We do this through an analysis of the UV excess, the blackbody evolution, late-time spectra and, when available, public IR data. We search for differences between the SLSNe~II that seem to require a central engine and those that may be powered through CSI alone. The data available to us are summarized in Sect. \ref{sec:sample}; the blackbody fits and UV/IR excesses are examined in Sect. \ref{sec:BB} and late-time spectra in Sect. \ref{sec:spex}. We discuss our results and their implications in Sect. \ref{sec:disco} and summarize our conclusions in Sect. \ref{sec:conc}.

Throughout this paper, we use the AB magnitude system \citep{okegunn83} unless otherwise stated, and we assume a $\Lambda$CDM cosmology with parameters $H_0 = 69.6$~km~s$^{-1}$~Mpc$^{-1}$, $\Omega_M = 0.286$ and $\Omega_\Lambda = 0.714$ \citep{bennett14}. Reported uncertainties correspond to 68\% ($1\sigma$) and upper limits to $3\sigma$. All absolute magnitudes have been $K$-corrected \citep{hogg02} using the first-order term of $K = -2.5 \mathrm{log}(1+z)$, which was shown by \citetalias{kangas22} to be a reasonable approximation in SLSNe II.

\section{Sample and data}
\label{sec:sample}

\subsection{Previously published data}

In this paper, we studied the late-time spectroscopic behavior and SED evolution of SLSNe~II in order to constrain the mass loss of their progenitor stars. We therefore used the sample of SLSNe~II in the literature with relevant data (i.e., late-time optical spectra and optical, NIR and/or NUV photometry) available to us. We included the following objects and data:
\begin{itemize}
    \item SN~2008es, the prototypical SLSN~II. We made use of the optical, NIR and UV photometry and optical spectra in \citet{gezari09} and \citet{miller09} and the late-time spectrum and photometry in \citet{kornpob19}.
    \item The events in \citetalias{kangas22}, with the exception of AT~2020yue, which was later classified as a TDE based on late-time data \citep{yao23}. For these SNe, we made use of all the data published in \citetalias{kangas22}, including the spectral sequences and optical and UV photometry. We additionally included late-time spectra of SN~2019cqc published by \citet{hebbar26}.
    \item SN~2023gpw \citep{kangas26}. For this SLSN II, we included the published optical, NIR and UV photometry and spectral sequence.
\end{itemize}
For details on the reduction of these data, we refer the reader to the papers cited above. For different targets, there is a wide range of data availability; not all targets have available late-time spectroscopy (discounting spectra with only host lines visible) or NUV coverage, and NIR follow-up is extremely scarce. When coverage is available, its time scale also varies. We summarize the sample and the available data -- the optical photometric coverage and the included (rest-frame) filters, the NUV coverage and the late-time spectroscopic coverage -- in Table~\ref{tab:coverage}.

\subsection{Infrared imaging}
\label{sec:unwise}

We additionally obtained the publicly available \textit{Wide-field Infrared Survey Explorer} (\textit{WISE}) data through the Unblurred Coadds of the \textit{WISE} Imaging \citep[unWISE;][]{lang14,meisner17a,meisner17b}.\footnote{\url{https://unwise.me/}} We downloaded all available imaging in the \textit{WISE} $W1$ and $W2$ bands (respectively, 3.4 and 4.6~$\mu$m in the observer frame).\footnote{\url{https://portal.nersc.gov/project/cosmo/temp/ameisner/}} 

\begin{figure}
\centering
\includegraphics[width=0.95\linewidth]{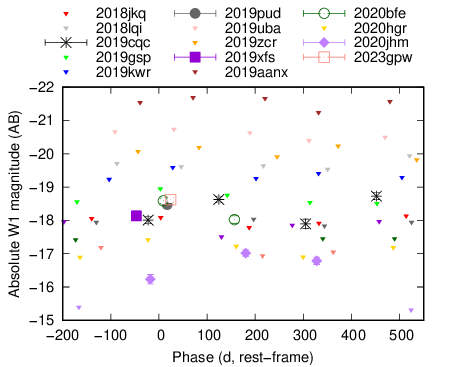}
\caption{Our unWISE $W1$ photometry of the sample SNe. For most events, only upper limits can be placed on the flux, but several of the closest SNe are detected. }
\label{fig:unwise}
\end{figure}

For each target, we median-combined all \textit{WISE} images available at least 200 rest-frame days before the $g$-band maximum of the SN in order to construct a deep template image with no contribution from the SN. This template was then subtracted from images taken after the explosion date using a modified version of the \texttt{ISIS}~2.2 package, and the methods described in \citet{alard98} and \citet{alard00}, within the \texttt{Image Reduction and Analysis Facility} \citep[\texttt{IRAF};][]{tody86}.\footnote{\url{https://iraf-community.github.io/}} After this, we performed aperture photometry on the subtracted images using \texttt{Starlink GAIA} \citep{currie14}, with an aperture radius of 6\arcsec, comparable to the full width half maximum (FWHM) of $6\farcs4$ in \textit{WISE} images, to obtain either measurements or (in most cases) upper limits for the brightness of the SNe in the \textit{WISE} bands until $\sim$600 rest-frame days after the $g$-band maximum of each target. In the case of SN~2008es, which exploded before the launch of \textit{WISE}, the earliest images are from MJD\,=\,55519, roughly 2.1~yr after the optical maximum, and \textit{WISE} data of this SN are not included in this analysis. The measurements and limits are listed in Table~\ref{tab:wisemags} in the Vega magnitude system. We show the absolute-magnitude light curves -- mostly upper limits -- of all objects in Fig.~\ref{fig:unwise}.

In total, we detect SNe~2019cqc, 2019pud, 2019xfs, 2020bfe, 2020jhm and 2023gpw in at least one epoch. All of these SNe, it must be noted, are located at redshifts of $z\lesssim0.12$ -- the implication is that for more distant events, even co-added unWISE images are not deep enough for detections. The absolute AB magnitudes reached by the detected SNe range from $-16$ to $-19$~mag in the $W1$ filter, while for the most distant events we can only place non-constraining upper limits around $-20$ to $-22$~mag. In the cases of SNe~2019pud, 2019xfs and 2023gpw, a detection can be made at only one relatively early epoch, ranging from $-47$~d for SN~2019xfs to +24~d for SN~2023gpw.

Out of the SNe detected at multiple epochs, the location of SN~2019cqc coincides with a static source, and we caution that this object may leave a residual in the difference image that affects the measured magnitudes. However, we do detect a source at the position of SN~2019cqc in each subtracted image; this corresponds to epochs between $-22$ and +634~d. In addition, SN~2020jhm is detected in three epochs between $-18$ and +328~d in $W1$ and between +180 and +524~d in $W2$. This SN is both the closest SN in the sample with a redshift of $z=0.057$, and the only one to be detected in $W2$. Finally, SN~2020bfe is detected at +9 and +157~d. We examine these detections compared to blackbody function fits in Sect. \ref{sec:excess}.

\section{Blackbody parameters and UV excess}
\label{sec:BB}

\subsection{Fitting procedure}
\label{sec:BBfits}

We performed blackbody fits to the UV-optical-IR SEDs. NUV magnitudes were used at each available epoch. Optical light curves were interpolated to these epochs where needed; outside the available NUV coverage and between sparsely placed NUV epochs, we used the interpolated optical magnitudes at intervals of roughly 10 days or at epochs when IR data were available. Interpolation of IR data was usually impossible due to their scarcity. Interpolation was done using a Gaussian process (GP) regression algorithm \citep{rasmussen06} via the Python-based {\tt george} package \citep{hodlr}; The widely-used Mat{\'e}rn-3/2 kernel was applied, with a 0.05~mag error floor added to measured magnitudes to avoid over-weighting individual points, and hyperparameters were not fixed. Blackbody fits to objects with temperatures above a few $\times 10^4$~K with optical data alone introduce severe systematic effects \citep{arcavi22}, but in our sample this is only the case at a few very early epochs, where the fit is highly uncertain in any case.

As most of the targets exhibit a UV excess \citep{miller09,kangas22,kangas26}, we ignored the $Swift$/UVOT $UVW2$ and $UVM2$ filters in all blackbody fits. In cases where the $UVW1$ filter, the reddest of the available NUV filters, also showed inconsistency with the blackbody, we repeated the fit ignoring this filter as well -- but this was not common. In the case of SN~2023gpw, it was reported by \citet{kangas26} that at early epochs (until +4~d), one blackbody is enough to describe the SED including the NUV. However, for internal consistency and because the (observed) $u$-band point slightly deviates from this single blackbody, we have also applied this method to SN~2023gpw, which results in somewhat different early blackbody parameters than reported in \citet{kangas26}. 

We also fitted a constant velocity to the early expansion of each blackbody radius. For this purpose, we included epochs up to the point where the photospheric expansion starts decelerating by eye. These velocities were then extrapolated to zero radius to approximate an explosion epoch. This estimate was compared to those obtained through light-curve modeling to look for clear discrepancies. For SN~2008es, we took the date estimated through the CSM+$^{56}$Ni model in \citet{inserra18}. For other events, we used the median explosion dates and their uncertainties from the best-fitting magnetar or CSI models presented in \citetalias{kangas22} and \citet{kangas26}.

\subsection{Observed excesses}
\label{sec:excess}

We show the blackbody fits to SEDs in the Appendix. Fig.~\ref{fig:BB} shows the cases where the blackbody fit is generally based on more than three bands, whereas Fig.~\ref{fig:BB2} shows the SNe where the blackbody fits are based on two or three (optical) bands; in most of these cases, furthermore, only limits are available in the IR. In Fig.~\ref{fig:BB2}, we only show the blackbody fit(s) relevant to the existence of an IR excess, i.e. coinciding with an IR epoch. Note that our blackbody fits were performed without \textit{WISE} bands.

An NUV excess has been reported previously in SNe~2008es \citep{miller09}, 2019uba, 2019zcr, 2020hgr, 2020jhm \citepalias{kangas22} and SN~2023gpw \citep{kangas26}. We additionally show SNe 2019cqc and 2019xfs, where $Swift$/UVOT measurements exist but are mostly insignificant ($<3\sigma$) and hence were not presented by \citetalias{kangas22}. In these two, we also see a deviation from a blackbody in the NUV at later epochs (for 2019cqc, starting at +41~d, and for 2019xfs, starting at +45~d), but apart from the one significant $UVW2$ point in the +45-day SED of SN~2019xfs, the deviations are within $\sim2\sigma$. Taken together, these observations do not prove an excess in these two SNe, but they do suggest it. We examine the UV excesses in more detail in Sect. \ref{sec:UVex}.

In the IR, on the other hand, we mostly see consistency with a blackbody. In SN~2019cqc, this is not the case either at early ($-24$~d) or late ($+120$~d) times. However, as was pointed out in Sect.~\ref{sec:unwise}, this SN occurred at a location coincident with a static source, which likely affects our \textit{WISE} subtractions and photometry. \citet{hebbar26} found signs of dust formation in late-time spectra of this SN, but at early times, before the optical peak, this should not produce an IR excess yet. In SN~2020bfe, meanwhile, the SED at +4~d is compatible with a blackbody from optical to IR, while at +152~d there seems to be an excess in the IR. We note, though, that in this case the optical blackbody fit is based on only three points, and may be unreliable.

Therefore, no clear IR excesses are seen at epochs where shorter-wavelength data are available, and in the other SNe with \textit{WISE} detections (2019pud, 2019xfs and 2023gpw) the IR SED is consistent with a blackbody. In a few SNe where only limits are available (SNe 2019uba, 2019zcr and 2020hgr), these limits are close to the blackbody fit and do not allow a large excess at $\sim3\,\mu$m. The same applies in SNe 2018jkq and 2018lqi, but as the blackbody fit is only based on two bands in these cases, we cannot say how constraining their IR limits are. 

\subsection{Blackbody evolution}
\label{sec:BBevo}

\begin{figure*}
\centering
\includegraphics[width=0.42\linewidth]{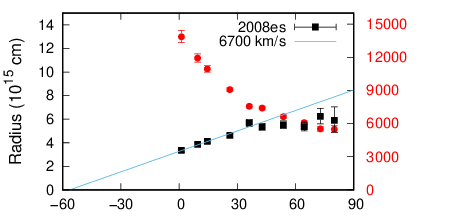}
\includegraphics[width=0.42\linewidth]{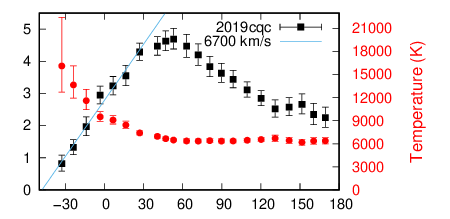}
\includegraphics[width=0.42\linewidth]{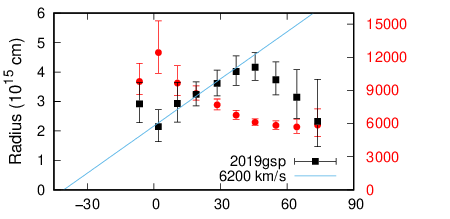}
\includegraphics[width=0.42\linewidth]{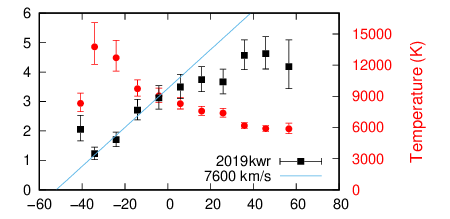}
\includegraphics[width=0.42\linewidth]{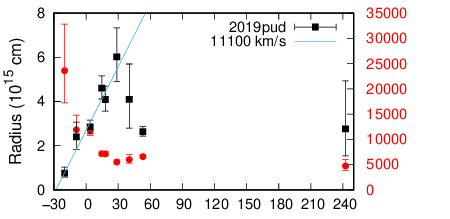}
\includegraphics[width=0.42\linewidth]{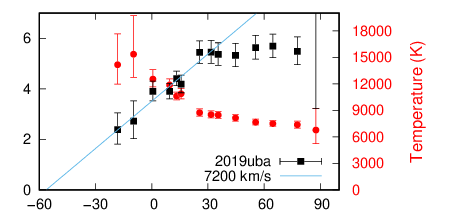}
\includegraphics[width=0.42\linewidth]{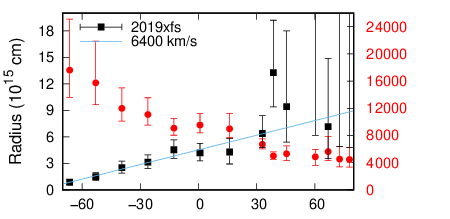}
\includegraphics[width=0.42\linewidth]{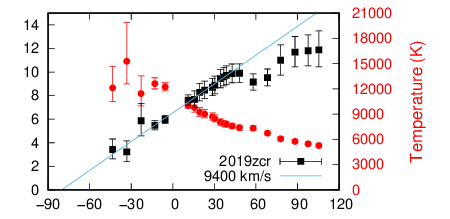}
\includegraphics[width=0.42\linewidth]{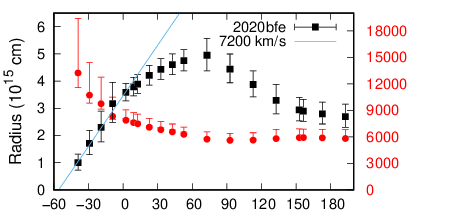}
\includegraphics[width=0.42\linewidth]{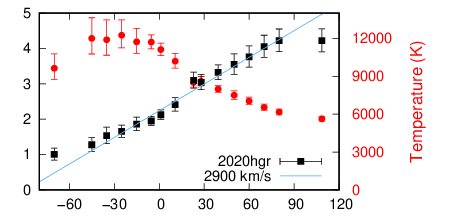}
\includegraphics[width=0.42\linewidth]{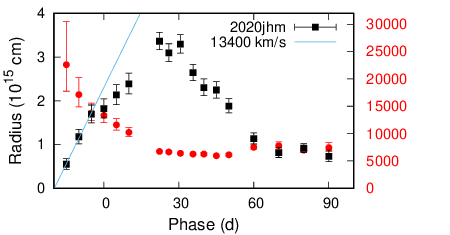}
\includegraphics[width=0.42\linewidth]{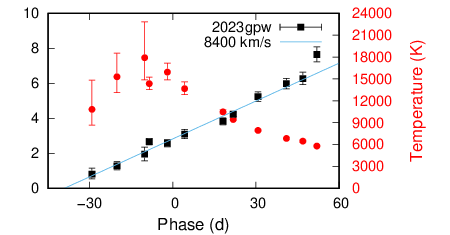}
\caption{Blackbody radius (black points) and temperature (red points) evolution and expansion velocity fits (blue lines) of the sample SNe with enough data.}
\label{fig:BBevol}
\end{figure*}

The temperatures of the objects in our sample typically reach between 10\,000 and 15\,000~K between the explosion and the optical peak, while in some cases, roughly 20\,000~K is reached in very early epochs (these are often based on only two or three data points, however). All objects decline to temperatures close to 6\,000~K (i.e., close to the hydrogen recombination temperature) a few months after peak. Meanwhile, the blackbody radius starts stalling and then, as eventually seen in some of the SNe, shrinking, as it recedes in velocity space after what looks like an initial constant-velocity phase where the outer ejecta is still optically thick. In several other objects the shrinking phase is not reached, though: SN~2023gpw becomes unobservable before its photosphere recedes past the outer layers, while SNe~2020hgr, 2019zcr, 2019uba, 2019kwr and 2008es only seem to start stalling during the multi-wavelength observations. For SN~2019xfs, huge uncertainties preclude any conclusions.

We summarize the features of the blackbody expansion of each SN in Table~\ref{tab:bbvels}. This includes comparisons between the expansion velocity $v_\mathrm{BB}$ and the spectroscopic velocity $v_\mathrm{H\alpha}$. The latter was estimated from the H$\alpha$ line for uniformity, as this was the only line reliably available in each target. We used the absorption minimum at early times (during the epochs used for the early velocity fit) when possible, or the FWHM of the emission if absorption was not clearly seen -- and if the H$\alpha$ line only becomes measurable later, we used the first possible epoch. We note that in the absence of an absorption trough, the blue edge of the emission is a better measure of the direct line-of-sight velocity of the outer ejecta, but is more difficult to measure robustly. However, as a consistency check, we estimated the blue edge velocities by eye as well, and found that the measured FWHM and the estimated blue edge velocity typically agree within $\sim$20 per cent (the FWHM overestimating the velocity by $\sim$10 per cent on average), which should therefore be treated as an additional systematic error on the FWHM velocities. We also compared the explosion dates estimated from the blackbody radius extrapolation and the best-fitting light-curve model. 

\begin{table*}
\centering
\caption{Blackbody radius expansion properties of each SN.}
\label{tab:bbvels}
\begin{tabular}{ccccccccc}
\hline
SN & $v_\mathrm{BB}$ & $v_\mathrm{H\alpha}$ & Type & Epoch & $t_\mathrm{expl,BB}$ & $t_\mathrm{expl,model}$ & Model & Model\\
  & (km s$^{-1}$) &  (km s$^{-1}$) & & (d) & (d) & (d) & & reference \\
\hline
2008es & $6700\pm730$ & $9300\pm580$ & FWHM & +45\,¤ & $-56.4\pm6.5$ & $-36.4$ & CSM$+^{56}$Ni ($s=0$) & 1 \\
2019cqc & $6740\pm400$ & $13050\pm460$ & Min & +29 & $-48.0\pm3.1$ & $-43.4\pm0.6$ & Magnetar$+^{56}$Ni & 2 \\
2019gsp & $6180\pm930$ & $8610\pm190$ & FWHM & +75\,¤ & $-40.6\pm6.8$ & $-27.7^{+1.6}_{-2.3}$ & CSM$+^{56}$Ni ($s=2$) & 2 \\
2019kwr & $7650\pm950$ & $6600\pm180$ & FWHM & +64\,¤ & $-52.1\pm7.2$ & $-52.9^{+1.0}_ {-1.7}$ & Magnetar$+^{56}$Ni & 2 \\
2019pud & $11100\pm1100$ & $12100\pm300$ & FWHM & +52\,¤ & $-28.1\pm3.2$ & $-29.4^{+1.4}_{-2.1}$ & CSM$+^{56}$Ni ($s=0$) & 2 \\
2019uba & $7200\pm1300$ & $4970\pm460$ & FWHM & +30 & $-56.3\pm10.0$ & $-33.4^{+1.7}_{-1.8}$ & CSM$+^{56}$Ni ($s=2$) & 2 \\
2019xfs & $6390\pm500$ & $8210\pm290$ & Min & +73 & $-82.0\pm7.8$ & $-81.8\pm0.4$ & Magnetar$+^{56}$Ni  & 2\\
2019zcr & $9430\pm390$ & $5680\pm310$ & FWHM & +46 & $-80.3\pm3.5$ & $-77.4^{+1.8}_{-2.0}$ & CSM$+^{56}$Ni ($s=0$) & 2\\
2020bfe & $7170\pm370$ & $7680\pm110$ & FWHM & +63\,¤ & $-55.9\pm3.3$ & $-64.1\pm1.2$ & Magnetar$+^{56}$Ni & 2\\
2020hgr & $2910\pm150$ & $6230\pm170$ & Min & +81 & $-89.0\pm4.7$ & $-96.2^{+1.5}_{-1.6}$ & CSM$+^{56}$Ni ($s=0$) & 2\\
2020jhm & $13440\pm740$ & $14200\pm700$ & FWHM & +33\,¤ & $-19.9\pm1.3$ & $-18.8\pm0.2$ & CSM$+^{56}$Ni ($s=2$)  & 2\\
2023gpw & $8350\pm510$ & $5500\pm300$ & FWHM & +47 & $-39.1\pm2.8$ & $-32.9\pm0.3$ & CSM ($s=0$) & 3\\
\hline
\end{tabular}
\tablefoot{Dates are relative to the rest-frame $g$-band peak of each target. H$\alpha$ velocities are measured during the phase of constant blackbody expansion if possible. Measurements taken from the first available spectrum after the blackbody velocity fit are marked with ¤. References correspond to: 1) \citet{inserra18}, 2) \citetalias{kangas22}, 3) \citet{kangas26}. }
\end{table*}

As reported in \citet{kangas26}, there is a discrepancy between $v_\mathrm{BB}$ and $v_\mathrm{H\alpha}$ in SN~2023gpw; namely, $v_\mathrm{BB}$ is considerably higher (by a factor of 1.5 or $\sim$3500~km~s$^{-1}$). Since the two velocities measure different things -- $v_\mathrm{BB}$ being related to transverse expansion and $v_\mathrm{H\alpha}$ along the line of sight -- asphericity in the shape of the photosphere could cause such a discrepancy. We therefore investigated if any other objects in the sample show signs of asymmetry in their blackbody behavior. In SNe~2019uba and 2019zcr, a similar velocity discrepancy exists. However, the significance of the discrepancy is only $1.6\sigma$ in SN~2019uba as the blackbody fit has large uncertainties. In SNe~2019kwr and 2019xfs the discrepancy is only a factor of 1.2 and 1.3, respectively, while in some cases (SNe~2008es, 2019cqc, 2019gsp and 2020hgr), $v_\mathrm{BB}$ is instead lower than $v_\mathrm{H\alpha}$ by about a factor of 1.5--2. Objects where the two velocities are consistent within $\sim$1$\sigma$ are SNe~2019pud, 2020bfe and 2020jhm. 

The blackbody-based extrapolated explosion date is based on an assumed constant velocity, but this is not always the case, especially if the first blackbody fit is done at a relatively late epoch. The photosphere may be already receding during the epochs when the velocity fit is performed, however linear its expansion, in which case $v_\mathrm{BB}$ does not directly measure the velocity of matter at the photosphere. This is most likely the case for SN~2008es, where the explosion date discrepancy is $\sim$20 days. The velocity required to reach the first blackbody radius close to the $g$-band peak, assuming $t_\mathrm{expl,model} = -36.4$~d, is $\sim$10\,000~km~s$^{-1}$, and for subsequent blackbody radii it declines linearly to reach $\sim$6\,000~km~s$^{-1}$ at $\sim$+100~d. Therefore $v_\mathrm{H\alpha}$ can originate well outside the photosphere after the earliest blackbody fit and the lower $v_\mathrm{BB}$ is easily explained. The same also applies to SN~2019gsp, where assuming $t_\mathrm{expl,model}$ instead of $t_\mathrm{expl,BB}$ results in velocities steadily declining from 8\,000 to 6\,000~km~s$^{-1}$. 

However, in SNe 2019cqc and 2020hgr, the date discrepancy is small and the first blackbody fit relatively early, suggesting that the velocity discrepancy needs a different explanation, such as a non-spherical blackbody resulting in an underestimated radius and hence $v_\mathrm{BB}$. Despite the slightly different early blackbody fitting for SN~2023gpw, $v_\mathrm{BB} > v_\mathrm{H\alpha}$ by about 3\,000~km~s$^{-1}$, and the first blackbody fit is also quite early; therefore the velocity discrepancy remains, and in this case H$\alpha$ \citep[and absorption lines at a similar velocity; see][]{kangas26} cannot simply come from a layer above a spherical, receding photosphere expanding at $\geq$8\,000~km~s$^{-1}$. As for SN~2019zcr, $t_\mathrm{expl,model}$ and $t_\mathrm{expl,BB}$ are consistent again, and the increase of the radius is close to linear for a long time. Here $v_\mathrm{BB} > v_\mathrm{H\alpha}$ by about 3\,700~km~s$^{-1}$, which could indeed measure bulk velocities in different directions. We discuss the implications further in Sect. \ref{sec:disco}.

\subsection{UV excess strength and mass loss}
\label{sec:UVex}

In order to quantify the UV excess in each SLSN~II, we subtracted the best blackbody fit from the SED. As shown in Fig.~\ref{fig:BB}, these fits were performed using specific luminosities ($L_\lambda$, in units of erg~s$^{-1}$~\AA$^{-1}$) at each wavelength, either without \textit{UVW2} and \textit{UVM2} points, or in some cases, without any NUV points, usually resulting in an excess in those bands. 

\begin{figure}
\centering
\includegraphics[width=0.92\linewidth]{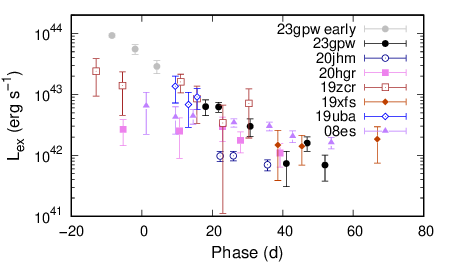}
\includegraphics[width=0.92\linewidth]{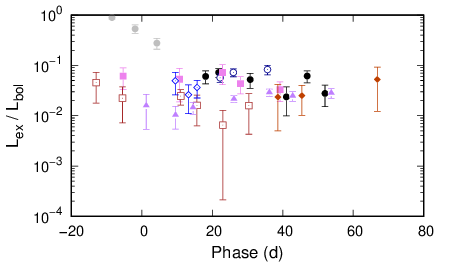}
\includegraphics[width=0.92\linewidth]{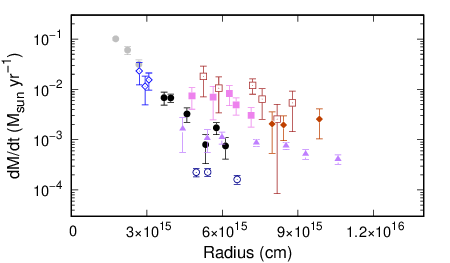}
\caption{First panel: UV excess luminosity $L_\mathrm{ex}$ as a function of epoch in each target. Second panel: $L_\mathrm{ex}$ as a fraction of the total luminosity $L_\mathrm{bol}$, consistently around a few per cent. Third panel: mass-loss rate estimates based on the UV excess as a function of radius from the center of the explosion, estimated through the model-based explosion epoch and the ejecta velocity. 
Epochs when the excess significance is $<1\sigma$ are omitted. Points marked in grey correspond to the early epochs of SN~2023gpw where one blackbody can reasonably fit the SED.}
\label{fig:excess_L}
\end{figure}

We then integrated $L_\lambda$ over the wavelength range of the \textit{UVW2} and \textit{UVM2} filters in epochs where both filters were available. Therefore, the resulting UV excess luminosity is from these observed NUV filters only and, in practice, a lower limit. The rest-frame spectral range covered by these filters is redshift-dependent, but in the observer frame we integrated from the blue edge (defined here as the bluest wavelength where the transmission efficiency is $\geq1\%$) of \textit{UVW2} to the peak transmission wavelength of \textit{UVW1} (where $L_\lambda$ is typically consistent with the blackbody), corresponding to the range of 1602.5 to 2612.5~\AA. The integration was trapezoidal, assuming the spectrum between the effective wavelength and blue edge of \textit{UVW2} is flat in each case. We show the resulting excess luminosity ($L_\mathrm{ex}$) light curves in the top panel of Fig.~\ref{fig:excess_L}. In many cases, though not always, the uncertainties of both the measured SED and the blackbody fit result in a low significance (1--3$\sigma$) for the excess, though it is apparent by eye compared to the best-fit function. The contribution of the UV excess to the total pseudo-bolometric luminosity is typically between 1 and 10 per cent, as shown in the second panel of the Figure. This fraction does not evolve significantly with time.

Assuming that the UV excess is powered by CSI and that the interaction luminosity mainly emerges in the UV \citep[see e.g.][]{dessart22}, we then estimated the mass loss required for it. This is, in practice, an underestimate, as in the absence of a central engine the emission from the CSI may also dominate in the optical, but it serves as a lower limit. We estimated the mass loss as $\frac{dM}{dt} = (2L_\mathrm{ex}/\epsilon) v_{\mathrm{CSM}}v^{-3}_{\mathrm{sh}}$ \citep[e.g.][]{smith17}. Here, $v_\mathrm{CSM}$ is the velocity of the CSM, $v_{\mathrm{sh}}$ is the velocity of the forward shock and $\epsilon$ is the efficiency of converting kinetic energy into radiation through the interaction, for which we assume $\epsilon=0.5$. For $v_\mathrm{CSM}$, we assumed 100~km~s$^{-1}$, typical for luminous blue variable (LBV) or yellow hypergiant (YHG) stars \citep[e.g.][]{smith14,smith17}.\footnote{In the 2008es-like SN~2024adtg (Pylypenko et al., in preparation), CSM velocities of this order are seen when such velocities can be resolved, supporting this choice.} Note that for pulsational pair instability, velocities can reach thousands of km~s$^{-1}$ \citep{woosley17}, in which case all mass-loss rates would be adjusted upwards by a factor of $\sim$10. For $v_{\mathrm{sh}}$ we used the higher of the $v_\mathrm{BB}$ and $v_\mathrm{H\alpha}$ values listed in Table \ref{tab:bbvels} to approximate the outer ejecta producing the excess through CSI. We estimate the radius corresponding to this CSM by using the best-fitting model explosion date and $v_{\mathrm{sh}}$ to find the location of the outer ejecta at each epoch. 

We show our mass-loss rate estimates in the third panel of Fig.~\ref{fig:excess_L}. These values mostly range from $\sim10^{-4}\, \mathrm{M_\odot}$~yr$^{-1}$ to $\sim10^{-2}\, \mathrm{M_\odot}$~yr$^{-1}$. The early epochs of SN~2023gpw show higher values, but as one blackbody reasonably reproduces the SED at these epochs, these values should not necessarily be taken at face value -- though they do also correspond to the shortest radii from the center of the explosion and thus the most confined, most recent mass loss. With $v_\mathrm{CSM} = 100$~km~s$^{-1}$, these radii (between 10$^{15}$ and 10$^{16}$~cm) correspond to mass lost in the several years to a few decades prior to the explosion. Given that there is, in most of the objects, a trend of increasing mass loss with decreasing radius (and thus closer in time to the explosion), it is conceivable that the early high values in SN~2023gpw simply continue that trend. Here the shock velocity is assumed to be constant; however, a decreasing velocity over time would increase mass-loss rates further out.

Assuming a constant mass loss rate and spherical CSM out from radius $R_0$, we can estimate the optical depth of the CSM to infinity by treating it as a series of shells with mass $dM$:
\begin{equation*}
    \tau \sim \int_{R_0}^{\infty} \frac{dM \kappa}{4 \pi R^2} = \frac{dM}{dt} \frac{\kappa}{4\pi}\int_{R_0}^{\infty} \frac{dR}{R^2 v_\mathrm{CSM}} = \frac{dM}{dt} \frac{\kappa}{4\pi R_0 v_\mathrm{CSM}} ~.
\end{equation*}
If we now take $R_0$ and $\frac{dM}{dt}$ to be the radius and corresponding mass loss rate at the first epoch with a clear UV excess, and assume $\kappa = 0.34$~cm$^{2}$~g$^{-1}$ and $v_\mathrm{CSM} = 100$~km~s$^{-1}$, we obtain a large range of optical depth estimates, roughly corresponding to the degree of variation in $\frac{dM}{dt}$ in Fig.~\ref{fig:excess_L}, between SN~2020jhm at $\tau \sim0.01$ and SNe~2019uba and 2019zcr at $\tau \sim1$. 

\section{Late-time spectra}
\label{sec:spex}

\begin{figure*}
\centering
\includegraphics[width=0.95\linewidth]{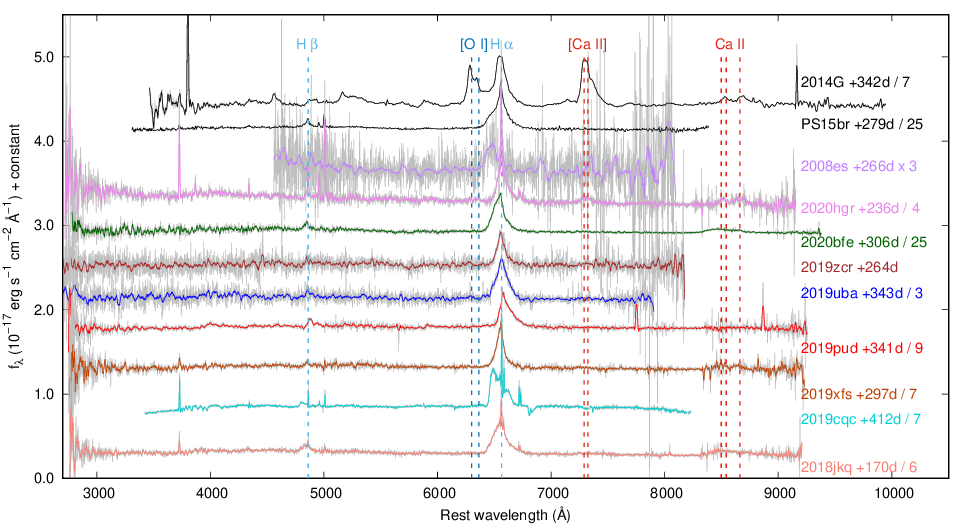}
\caption{Late-time spectra of our sample, compared to that of a normal-luminosity SN~II, 2014G \citep{terreran16}, and PS15br, a SLSN~II with a multi-component H$\alpha$ line \citep{inserra18}. H$\alpha$ is the only strong line in all sample objects, with much weaker H$\beta$ and Ca~{\sc ii}; forbidden lines can only be seen in SN~2020hgr, much weaker than in SN~2014G. Adapted from \citetalias{kangas22}, with the addition of the late-time spectra of SNe~2008es \citep{kornpob19} and 2019cqc \citep{hebbar26}. Savitzky-Golay smoothing has been applied; original un-smoothed spectra are shown in grey.}
\label{fig:latetime}
\end{figure*}

We show the late-time spectra available for our sample in Fig.~\ref{fig:latetime}. As reported by \citet{inserra18} and \citetalias{kangas22}, the late-time spectra of SLSNe~II are dominated by Balmer series emission, with little to no emission in forbidden lines of [O~{\sc i}]\,$\lambda\lambda6300,6364$ or [Ca~{\sc ii}]\,$\lambda\lambda7291,7324$, which are typical to nebular spectra of normal SNe~II, with fluxes on the same order as that of H$\alpha$. The spectrum of SN~2008es at +266~d \citep{kornpob19} is very noisy, but strong forbidden lines are absent there as well. SN~2020hgr is the only object in which a [Ca~{\sc ii}] doublet is clearly present, but even there it is weaker by far than H$\alpha$. In some cases, Ca~{\sc ii} IR triplet emission is visible but relatively weak. This, combined with a red-wing occultation visible in some objects, led \citetalias{kangas22} to suggest that the spectra are not truly nebular even at +200 or +300 days, as the density at the emitting region is still high enough to collisionally de-excite the transitions that would otherwise emit the normally-observed forbidden lines. We also note that \citet{inserra18} found the late-time Balmer emission of PS15br and SN~2013hx to have multiple components, which is not the case with the spectra in \citetalias{kangas22}. However, \citet{hebbar26} showed that the spectrum of SN~2019cqc at +412~d shows a triple-Gaussian profile suggesting a disk-like CSM, which loses its red component in their last spectrum at +488~d, possibly due to dust formation. Here, we further examined our late-time spectra in order to find other constraints on the ejecta or CSM structure. 

\begin{figure}
\centering
\includegraphics[width=0.9\linewidth]{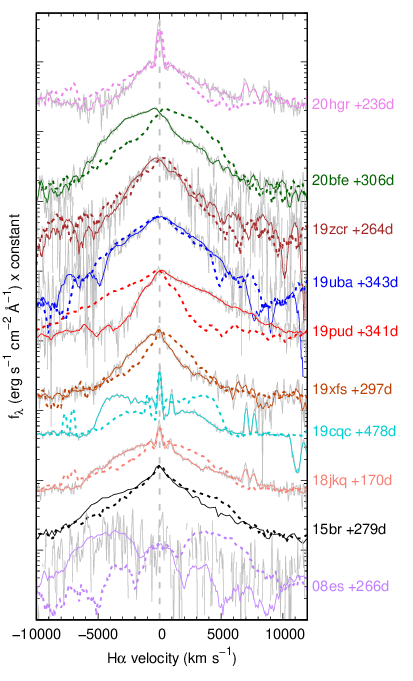}
\caption{H$\alpha$ profiles of our targets at late times, with the multi-component PS15br \citep{inserra18} as comparison. Dashed lines correspond to a line profile mirrored about the zero velocity. Asymmetric line profiles are common. Out of the most luminous events, SNe~2019zcr and 2019uba show a symmetric profile, while the profile of SN~2008es is very noisy but blueshifted. Host-galaxy lines of H$\alpha$, [N~{\sc ii}]\,$\lambda\lambda6548,6584$ and [S~{\sc ii}]\,$\lambda\lambda6717,6731$ are visible in SNe~2018jkq and 2020hgr. Savitzky-Golay smoothing has been applied; original un-smoothed spectra are shown in grey.}
\label{fig:hasymm}
\end{figure}

We started by examining the H$\alpha$ profiles of our targets; these are shown in Fig.~\ref{fig:hasymm} along with mirrored versions to study asymmetries in the profile. We see from this figure that the line profiles of SNe~2019zcr and 2019uba, two of the three most luminous SLSNe~II for which we have late-time spectra available, are the only ones where the profile is symmetric and roughly exponential, resembling electron scattering profiles \citep[see][]{huang18}. In SN~2020hgr, despite the weak forbidden [Ca~{\sc ii}], we still see a P Cygni profile with shallow absorption, and in SN~2019pud, an excess of emission in the red wing. A blue shoulder as prominent as that seen in PS15br \citep{inserra18} is not seen, but in SNe~2018jkq, 2020bfe and 2019xfs, the profile is clearly skewed to the blue. In each case the peak of the line is only slightly blueshifted or at zero velocity. Finally, in SN~2008es, we see a blueshift even higher than in the blue peak of SN~2019cqc, but the spectrum is too noisy to make out multiple components with any confidence.

We estimated the H$\alpha$/H$\beta$ line ratio of each spectrum by fitting a Gaussian profile to each line, ignoring any narrow components with the assumption that these are host-dominated (as is clearly the case in SN~2020hgr, for example). From this fit, we also extract the FWHM of the H$\alpha$ line. For the complicated H$\alpha$ profile in SN~2019cqc, we instead use a triple-Gaussian fit, and in the extremely noisy spectrum of SN~2008es, we only fit the H$\alpha$ line. We show these values in Table~\ref{tab:hahb}. In general, the H$\alpha$/H$\beta$ ratio is on the order of 10 or even 20, with the lowest measured value being $\sim$8 -- though some H$\beta$ lines are weak and noisy, resulting in large uncertainties in the ratio. In any case, it is clear that the theoretical value of 2.86 in optically thin ionized gas is not seen. This can in some cases indicate dust formation, especially if the red wing of the line is occulted (such as in SNe~2019cqc and 2020bfe), but reproducing such high ratios through extinction alone, assuming the \citet{gordon23} extinction law, requires improbably high values of $A_V \approx 4$~mag. Instead, high density and collisional excitation in the emitting region can increase the ratio, as the $n=3$ level is relatively more affected by it than the $n=4$ level \citep{agnagn}. This is especially the case for the objects without red-wing occultation, such as SNe~2019zcr and 2019uba. 

If the emitting region responsible for the late-time features is optically thick instead of thin, we can estimate an upper limit for a spherical late-time photosphere by using the estimated time since explosion and the velocity of the H$\alpha$ line (in the case of SN~2019cqc, using the blue peak velocity of $\sim$4000~km~s$^{-1}$ instead). For the explosion time, we again used the epoch estimated through the best-fitting light curve model. The resulting radii range from $\sim10^{16}$ to $\sim2\times10^{16}$~cm. From Fig.~\ref{fig:BBevol}, we can tell that the photosphere in most cases starts stalling or shrinking well before it reaches such values, though in the case of SN~2019xfs the errors are large and nothing certain can be said. SN~2019zcr reaches $>10^{16}$~cm before stalling, but even the photosphere of this SN is likely to shrink below such radii in the intervening $\sim150$~days instead of stalling indefinitely. Therefore, if the late-time spectra represent emission from a photosphere, it is unlikely to be spherically symmetric and could be, for example, a local photosphere around an asymmetric interaction region \citep[see][where a similar test was applied]{reynolds25,reynolds25b}. The filling factor of the CSM can be estimated as the ratio of the surface area of the emitting region implied by the line width and that of the blackbody fit. We have no blackbody fits at the relevant epochs, but we can set upper limits for the radius based on the last fit, if the radius is no longer increasing at that time. For example, for SN~2020bfe the filling factor $f$ would be $f \lesssim (2.7\times15\,\mathrm{cm} / 1.71\times16\,\mathrm{cm})^2 \approx 0.025$. A similar estimate is possible for SNe~2019pud ($f\lesssim0.024$), 2019cqc ($f\lesssim0.020$) and 2008es ($f\lesssim0.13$) -- all these values indicate a severe deviation from spherical symmetry and/or clumpiness. 

\begin{table}
\centering
\caption{Results of our H$\alpha$ and H$\beta$ fits at late times.}
\label{tab:hahb}
\begin{tabular}{cccc}
\hline
SN & Epoch & H$\alpha$/H$\beta$ & FWHM(H$\alpha$) \\
 & (d) & &  (km s$^{-1}$) \\
\hline
2020hgr & +236 & $16.5\pm6.2$ & $4630\pm150$ \\
2020bfe & +306 & $13.6\pm1.6$ & $5340\pm90$ \\
2019zcr & +264 & $11.3\pm3.2$ & $3870\pm170$ \\
2019xfs & +297 & $20.0\pm3.8$ & $3620\pm90$ \\
2019uba & +343 & $10.1\pm1.9$ & $5770\pm190$ \\
2019pud & +341 & $11.9\pm1.4$ & $5360\pm160$ \\
2019cqc & +412 & $22.5\pm3.9$ & - \\
2018jkq & +170 & $8.3\pm0.7$ & $6470\pm110$ \\
2008es & +266 & - & $6180\pm980$ \\
\hline
\end{tabular}
\tablefoot{Epochs are relative to the rest-frame $g$-band peak of each target. Note that the profile of SN~2019cqc has multiple components, and the single-Gaussian FWHM is omitted, while we omit the H$\beta$ fit for SN~2008es due to a high level of noise.}
\end{table}

\section{Discussion}
\label{sec:disco}

\subsection{Signs of asymmetry}

We first discuss the CSM configuration. \citet{nagao25} have suggested that, based on polarimetry, asymmetric CSM is common in luminous, interacting SNe~II, and though polarimetry of SLSNe~II is very scarce -- only two reported measurements, both suggesting asphericity \citep{inserra18,kangas26} -- SLSNe~II seem to be no exception. Indications for asymmetric CSM were already seen by \citet{inserra18}, whose targets, PS15br and SN~2013hx, both showed a multi-component late-time H$\alpha$ profile \citep[though we note that a blue shoulder similar to that in PS15br could also be produced without asymmetry if the profile is shaped by electron scattering and the red side is occulted; see][]{taddia20}. \citet{hebbar26} also note an asymmetric, multi-component profile in SN~2019cqc, which they interpret as a disk- or torus-shaped CSM. While this kind of profile is not seen in other events in the sample, we do see other phenomena that disfavor spherical symmetry.

The late-time spectra of all SNe in the sample are dominated by Balmer lines with high H$\alpha$/H$\beta$ ratios (on the order of 10; see Table~\ref{tab:hahb}) and lack prominent forbidden lines (Fig.~\ref{fig:latetime}). Only in SN~2020hgr can one discern any forbidden [Ca~{\sc ii}] emission, and it is very weak compared to those typical to nebular SNe~II. Such a non-nebular, optically thick spectrum indicates a relatively high density at the emitting region, causing both collisional excitation leading to high Balmer decrements \citep{agnagn}, and de-excitation of energy levels that might otherwise produce forbidden lines. At such late times, on the order of a year after the explosion, optically thick Balmer-dominated emission is common in SNe IIn \citep[see, e.g.,][]{fransson14,taddia20} and likely originates at an interaction photosphere, though the shape of the profile can vary with the CSM configuration.

As shown in Sect.~\ref{sec:spex}, if one assumes such a photosphere dominates the spectrum, the age of the SN and the velocity of the Balmer lines at late times result in radii that are excluded by our spherical blackbody SED fits. Of the objects where a comparison is possible, only SN~2019zcr actually reaches this estimated photospheric radius, and is technically compatible with a spherical interaction photosphere (filling factor $f\leq1$). By assuming that the width of the H$\alpha$ line is dominated by ejecta velocity, we obtain $f\lesssim0.09$ for SN~2019uba, but if it is instead dominated by electron scattering, SN~2019uba may also be spherical at late times. In other cases, however (SNe~2019cqc, 2019pud, 2020bfe, 2020hgr and possibly 2019xfs), the photosphere may instead be located around an interaction region in a highly asymmetric ($f$ of a few per cent), for example disk-like, CSM engulfed by the ejecta, as suggested for SN~2021irp by \citet{reynolds25,reynolds25b}. The interaction would ionize the surrounding H-rich matter, making it more opaque than the rest of the ejecta. If we look at such a system at an intermediate angle, we can observe relatively high line-of-sight spectral velocities without a large blackbody being necessary and without narrow lines. In SN~2020hgr, where weak forbidden line emission is seen, this can be the result of having a line of sight to a relatively more luminous optically thin ejecta as well. We lack a late-time spectrum for some SLSNe~II, but for CSS121015 \citep{benetti14} and SNe 2019kwr, 2020jhm \citepalias{kangas22}, 2023gpw \citep{kangas26} and 2024adtg (Pylypenko et al. in preparation), the last published spectrum at a few hundred days seems to be dominated by the host galaxy, perhaps suggesting that strong interaction has ceased by this time.

From the SED fits at early times, we can also extract the expansion velocity of the blackbody radius. If a spherical ejecta has a sharp outer edge and is almost entirely optically thick, this velocity corresponds to the real photospheric velocity. The typical scenario for a SLSN~II is one where the very early spectrum simply shows a blue continuum \citep[with possible flash emission features, such as in SN~2023gpw;][]{kangas26}, followed by the emergence of H$\alpha$ some time after the optical peak. At this point, the photosphere may thus indeed be located at the outer edge of the ejecta, with only a small amount of hydrogen above it. If the broad H$\alpha$ emerges as the photosphere recedes into the H-rich outer ejecta, the velocity of this line will, momentarily, roughly match the photospheric velocity. Later on, as the photosphere recedes further, the outer ejecta layers will be more and more detached from it and the photospheric velocity will be lower than the velocity of the outer layers. Several SNe in the sample can be explained through this simple scenario. 

This is not always true, however. In SNe~2019cqc and 2020hgr, we already see a blackbody expansion velocity clearly slower than the spectroscopic bulk velocity at early times (see Table~\ref{tab:bbvels}). Here, we can again invoke the overrun-disk CSM scenario: if the photosphere is dominated by the ejecta around the interaction region, its radius and hence the expansion velocity can be underestimated when assuming a spherical blackbody. In the case of SN~2020hgr, the first blackbody radius measurement seems to be an outlier; taking it at face value, it requires a velocity of $\sim$4\,500~km~s$^{-1}$, rather than the following 2\,900~km~s$^{-1}$. Here, we may even see the transformation of an initially roughly spherical photosphere into a torus as it gradually becomes dominated by the ionized region around the interaction. However, this epoch is based on only two points and is not particularly reliable. These two SNe are exceptions, but the late-time spectra suggest asymmetries in other events as well. It is possible that this seeming discrepancy is the result of varying viewing angles (which would change how much assumed sphericity affects the radius) and varying CSM/ejecta masses (a more massive CSM would have a larger effect on the shape of the photosphere). 

In some cases the expansion of the (spherical-equivalent) blackbody is instead faster than the spectral line velocities. These lines by definition originate outside the photosphere, but the two velocities do not measure the same thing. As discussed by \citet{kangas26}, if a non-spherical photosphere expands faster in the transverse direction than in the line-of-sight direction, the spectroscopic velocity can be considerably lower than the blackbody expansion. In addition to SN~2023gpw, this behavior is seen in SNe~2019zcr and 2019uba. Interestingly, all three objects have an energy budget that seems to require a central engine. If an engine launches bipolar outflows or jets, the ejecta itself may also be asymmetric in addition to, or instead of, the CSM. We also note, however, that SN~2008es does not show such a discrepancy, and its blackbody evolution can be explained through a receding photosphere. Instead, its late-time spectrum shows signs of asymmetric CSM with a filling factor $f\sim0.1$. We point out that, in order for a neutrino-driven explosion to be able to convert $10^{51}$~erg (about half of the available kinetic energy) into radiation, as reported for SN~2008es by \citet{miller09}, the filling factor of the CSM where the conversion occurs must be $f\gtrsim0.5$ -- therefore, such asymmetry provides further support for the central-engine power source in SN~2008es.

\subsection{Other CSM properties} 

A symmetric, exponential line profile is expected if the H$\alpha$ emission is strongly affected by electron scattering in extended CSM \citep{huang18}. \citet{taddia20} showed that a multi-component line seen through a medium increasingly optically thick to electron scattering becomes more symmetric, losing any red-wing attenuation and weak narrow components and developing broad wings. This scenario was suggested by \citetalias{kangas22} at earlier epochs for the three most luminous SLSNe~II in their sample, and may still be the case at late times. \citet{taddia20} also note that the optical depth to electron scattering $\tau_e$ in ionized CSM depends on the shock velocity as $\tau_e \propto v_\mathrm{shock}^{3}$; therefore, if $\tau_e$ is still high at late times, the shock must not have slowed down much. At the same time, the CSM must not have been wholly shocked yet, and strong narrow lines (whose flux is dependent on $\rho^2$) need to be avoided, implying an extended but not dense CSM. In these objects it is also possible that the CSM is spherically symmetric. The blackbody expansion is faster than the observed spectroscopic velocity in both cases, but this can be explained if the H$\alpha$ profile does not originate in the ejecta but in the CSM, rather than through any asymmetry. 

SLSNe~II in our sample, with NUV data of high enough quality to test it, show an excess at these wavelengths. \citetalias{kangas22} suggested its origin to be an optically thin, outer CSM, where continuing interaction with tenuous material produces the UV emission; a similar scenario was suggested for SN~2023ixf by \citet{bostroem24}. However, as shown in Fig.~\ref{fig:excess_L}, the excess luminosity seems to co-evolve with the pseudo-bolometric (mostly optical) luminosity, staying roughly between $L_\mathrm{bol} / 10$ and $L_\mathrm{bol} / 100$ depending on the SN; this ratio does not evolve much over time. This implies that the origins of the excess and the optical luminosity are linked, making such a "secondary" interaction component less likely. If the lack of narrow lines and signs of asymmetry discussed above are due to an aspherical CSM, overrun by the ejecta and heating it from within, this same interaction should also be responsible for the UV excess, which would emerge from the interaction photosphere. 

In models of interacting SNe~II but ignoring absorption and emission by CSM \citep{dessart22}, UV continuum flux from the cool dense shell between the forward and reverse shocks dominates the interaction output. In Sect. \ref{sec:UVex} we estimated a lower limit for the mass-loss rate responsible for the excess, assuming the luminosity of the interaction is mainly released in the UV. Even these rates are somewhat high -- between $\sim10^{-4}\, \mathrm{M_\odot}$~yr$^{-1}$ and $\sim0.03\, \mathrm{M_\odot}$~yr$^{-1}$ -- and if the optical luminosity is powered by the same mechanism, they become between one and two orders of magnitude higher. The estimated optical depth based only on the UV-excess mass loss rate is close to $\tau\sim1$ in SNe~2019uba and 2019zcr, which indicates that indeed a large fraction of the luminosity should be reprocessed into the optical, and the true mass loss rate must be higher.

In an upcoming paper (Pylypenko et al., in preparation), we examine SN~2024adtg, a spectroscopically 2008es-like SLSN~II with medium-resolution spectra from the X-Shooter instrument on the Very Large Telescope. These spectra show narrow absorption lines from CSM moving at $\sim300$~km~s$^{-1}$; this is similar to the velocities seen in LBV eruptions \citep{smith17}. CSM velocities of this order are realistic in other 2008es-like SNe, if not all SLSNe~II. The mass-loss rates estimated above are also similar to those associated with LBVs. However, even higher mass-loss rates are required if the optical luminosity is dominated by interaction as well: \citet{hebbar26} estimate a mass-loss rate of 0.2--0.3\, $\mathrm{M_\odot}$~yr$^{-1}$ for SN~2019cqc through analytical light-curve modeling, and \citet{kangas26} and Pylypenko et al. obtain a similar number for SNe~2023gpw and 2024adtg, respectively, from their bolometric luminosities, when assuming they are dominated by CSI. These numbers are consistent with other objects in this sample, such as SN~2019zcr, if one divides the UV-estimated mass loss rate by the $L_{ex}/L_{bol}$ ratio (Fig.~\ref{fig:excess_L}). 

\textit{WISE} detections and non-detections of SLSNe~II can in some cases be used to place constraints on any NIR emission from pre-existing circumstellar dust that might be causing an IR echo. However, as shown in Figs.~\ref{fig:BB} and~\ref{fig:BB2}, the detections we obtain almost always fall neatly on the blackbody function fitted to the (UV+)optical SED at the epochs when both are available, and limits at such epochs also tend to be close to the blackbody, implying no IR echoes at these times. For SNe at higher redshifts, the limits generally become less constraining (see Fig.~\ref{fig:unwise}). In SN~2019cqc, \citet{hebbar26} suggest the formation of dust at late times based on the Balmer line profiles; we see a supposed excess at both $-24$ and +120~d, but both epochs are earlier than when the dust formation is suggested to take place ($\gtrsim+400$~d), and as stated in Sect.~\ref{sec:unwise}, the photometry may be untrustworthy because of a residual from a static source. We may see a real IR excess in the SED of SN~2020bfe at +152~d, but even this is uncertain, as the blackbody fit is based on only three points and the excess is not large. 

The only high-confidence IR excess or echo in this sample is in SN~2020jhm, the nearest of the events, which is detected in \textit{W1} and \textit{W2} at +180 and +328~d and in \textit{W2} even at +524~d. By +180~d, the SN was no longer detected in the optical, but one can estimate that the $r$-band magnitude should be $\gtrsim22$~mag, resulting in an $r-W1$ color $\gtrsim2.1$~mag (AB system). As a comparison, SN~2021irp \citep{reynolds25} at $\sim+200$~d had both a clear IR excess and $r-W1$ between 0.6 and 1.3~mag based on \textit{WISE} photometry before and after this epoch, i.e., bluer than SN~2020jhm at +180~d; by +313~d SN~2021irp had reached $r-W1 \approx 3.6$. This suggests a strong IR excess comparable to that in SN~2021irp, possibly an echo from pre-existing dust, in SN~2020jhm, which seems to have faded a year later. As SN~2020jhm is spectroscopically unique in the \citetalias{kangas22} sample, this does not necessarily imply the presence of similar echoes in other objects, and none of our \textit{W1} upper limits can exclude an echo of similar absolute magnitude in the more distant SLSNe~II.

\subsection{Diversity within SLSNe~II}

SLSNe~II are not a homogeneous class. From the sample studied by \citetalias{kangas22}, it was noted that the sample seemed to divide into those SNe spectroscopically similar to SN~2008es in the photospheric phase -- also constituting the most luminous SLSNe II -- and the others, mostly similar to less luminous SNe II. The 2008es-like group currently consists of six objects: SN~2008es itself \citep{gezari09, miller09}, SNe 2019uba, 2019zcr and 2019aanx from the \citetalias{kangas22} sample, CSS~121015 \citep{benetti14} and SN~2024adtg, which is examined and compared to this subclass in an upcoming paper (Pylypenko et al. in preparation).\footnote{Spectroscopically, PS15br was also grouped in with the 2008es-like SNe by \citetalias{kangas22}, but its much fainter light curve bears no resemblance to the other events in the group, its late-time H$\alpha$ line shows multiple components, and its progenitor star may have been unrelated; we therefore do not consider it 2008es-like here.} This group is characterized by high emitted energies, on the order of (1--3)$\times 10^{51}$~erg, that seem to be out of reach for the neutrino-driven SN mechanism, even if most or all of their kinetic energy is converted into emission through CSI, and by star-forming dwarf host galaxies that resemble those of SLSNe~I \citep{schulze21,kangas22}. 

While the late-time spectra are all dominated by Balmer lines, the H$\alpha$ line profiles are far from uniform. Interestingly, two of the 2008es-like SNe that seem to require a central engine show the only symmetric H$\alpha$ profiles (Fig.~\ref{fig:hasymm}), likely shaped by electron scattering in optically thick CSM, while SN~2008es itself does not, instead exhibiting a heavily blueshifted profile, indicating that even within this group the CSM configurations can be diverse. SN~2008es reached an optical absolute magnitude of $\sim-15$~mag around +250~d \citep{kornpob19}, while SNe~2019zcr and 2019uba remained at a more luminous level around $-19$~mag at similar epochs \citepalias{kangas22}. From limits available for CSS121015 \citep{benetti14} and SN~2024adtg (Pylypenko et al. in preparation), these two SNe were also considerably fainter, $>-17.5$~mag, around +300~d. This diversity in late-time luminosity and spectral profiles can be explained simultaneously with a different extent of the CSM: SNe~2019uba and 2019zcr are still strongly interacting at late times with CSM extending to $\gtrsim10^{16}$~cm, giving them a boost in luminosity, and hiding the underlying H$\alpha$ profile behind a CSM still optically thick to electron scattering. Meanwhile, in the other objects, the CSM is less extended and has been swept up by these epochs.

\begin{figure}
\centering
\includegraphics[width=0.9\linewidth]{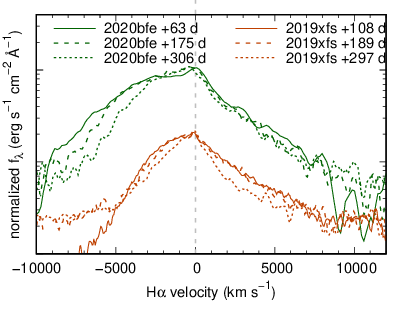}
\caption{Evolution of the blue-skewed H$\alpha$ profiles of SNe~2019xfs and 2020bfe. Continua have been subtracted to match the last epoch and the peaks of the lines are normalized to the last epoch.}
\label{fig:redwing}
\end{figure}

As for the other objects, while SN~2019pud is heavily skewed toward the red, SN~2020hgr still shows some absorption at +236~d, and other SNe are skewed to the blue. Clear signs of multiple components are seen in SN~2019cqc \citep{hebbar26}, PS15br and SN~2013hx \citep[see][as the latter's late-time FORS2 spectrum is not available on WISeREP\footnote{\url{https://www.wiserep.org/}}]{inserra18}. A red-side degradation in the line profile can indicate dust formation; this is clearly seen in SNe~2008es and 2019cqc, where dust formation was indeed suggested \citep{kornpob19,hebbar26}. In the case of SN~2008es, a NIR excess corresponding to $\sim1500$~K dust was also reported. However, in SN~2019pud the red side is instead stronger than the blue. We have no earlier spectra with a broad H$\alpha$ line to compare to in SN~2018jkq, but in Fig.~\ref{fig:redwing} we show the profile evolution of the remaining blue-skewed objects, SNe~2019xfs and 2020bfe. In both cases the line becomes narrower at late times. Very little change is observed on the red side of the profile in SN 2020bfe, but some red-side degradation can be seen in SN~2019xfs. Dust formation therefore is not uncommon in the sample, but neither is it ubiquitously seen in our spectra.

The luminous, broad-lined SNe studied by \citet{nagao25} are technically not SLSNe by the definition of the peak magnitude ($M < -20$~mag), but such magnitude limits for SLSNe~II are arbitrary, being originally based on SLSNe~I. A large fraction of SLSNe~II likely emerge from the same continuous progenitor population, with different late-time mass-loss histories and viewing angles creating the combination of similarity and diversity in spectra. Interaction with a binary companion naturally produces CSM disks, and is an obvious candidate mechanism for producing a continuum of disk properties through a variety of different binary configurations. The most energetic SLSNe~II, however, may require a fundamentally different progenitor population, as they seem to favor different host galaxies \citepalias{kangas22}, but the small sample makes this unclear. The situation is complicated by SN~2023gpw, which also seems to require a central engine, whose H$\alpha$ line profile is more reminiscent of the non-2008es-like subgroup, but which is, light-curve-wise, so far unique \citep{kangas26}; and by SN~2020jhm, which is not spectroscopically either 2008es-like or similar to a normal SN~II, resembling SN~1988Z \citep[e.g.,][]{turatto93} without a narrow emission component more than anything else \citepalias{kangas22}, and shows the fastest light-curve evolution and the highest line velocities in the sample. 

It is possible that the SLSNe~II requiring central engines are related to SLSNe~I through more than just their host galaxies. \citet{benetti14} noted a spectroscopic similarity between CSS121015 and some SLSNe~I, and suggested a similarity in their ejecta structures as well. In our upcoming paper (Pylypenko et al. in preparation), a similarity between spectral lines in SLSNe~I and 2008es-like SNe in general is also pointed out. Furthermore, in a subset of SLSNe~I, a broad H$\alpha$ line attributed to CSI has been observed, at least at late epochs \citep{yan15,yan17,pursiainen22} -- though the spectra of these objects never become CSI-dominated in the way the spectra of SLSNe~II do, it is possible that a continuum exists from normal SLSNe~I through interacting SLSNe~I to 2008es-like SLSNe~II, involving increasingly extended CSM and stronger interaction. However, the light curves of interacting SLSNe~I are bumpy, unlike those of SLSNe~II, which would also require the CSM distribution to be smoother in the latter case rather than comprised of multiple shells. The progenitor would also need to be stripped while simultaneously retaining a H-rich CSM. As \citet{yan17} point out, no helium lines were detected in their (small) sample, presenting the question of where the helium has gone if the hydrogen still remains. The source of a H-rich CSM around a H/He-poor progenitor star could instead be a donor in a close binary system \citep{humina26}; in the case of 2008es-like SNe, the two stars could even be in a common-envelope stage when the primary explodes, providing a smooth, extended but not very dense CSM. On the other hand, helium lines are seen in 2008es-like SNe, which complicates this simple picture. More work is required to better understand this connection.

A pair instability origin has also been suggested for some H-rich SNe. \citet{hiramatsu26} examined SN~2023vbw, which peaked at $M_r = -19.2$~mag after a rise of 190 days and spectroscopically resembled a normal SN~II at peak. SN~2023vbw was argued to be a promising pair-instability SN (PISN) candidate based on the slow rise and high required explosion energy and ejecta mass. However, the light curve of this event resembles that of iPTFH14hls \citep{arcavi17}, which was suggested to be powered by CSI involving multiple CSM shells, possibly from \textit{pulsational} pair instability. Clearly, one must consider whether some SLSNe~II in the aforementioned subgroups could be (pulsational) PISNe. However, all events in the \citetalias{kangas22} sample were found incompatible with the $^{56}$Ni decay model, their light curves are much faster to evolve than those of SN~2023vbw or SN~2018ibb \citep[a H-poor PISN candidate; see][]{schulze24}, and there are no clear iPTF14hls-like bumps from CSM shells in the light curves. It is likely that the sample considered here is unrelated to these events.

\section{Conclusions}
\label{sec:conc}

We have examined the blackbody evolution, late-time spectra and publicly available \textit{WISE} rest-frame NIR data of a sample of 15 SLSNe~II -- i.e., H-rich SNe more luminous than $-20$~mag at peak and dominated by broad rather than narrow (hydrogen) emission lines -- with the goal of constraining the properties of the CSM around their progenitor stars. Not all of these methods could be applied to the entire sample because of a lack of available data, demonstrating the need to obtain consistent, extensive data sets for these events. Our conclusions are as follows:
\begin{itemize}
    \item Though polarimetry exists only for two SLSNe~II, it is also possible to see other signs of asymmetry in their blackbody evolution, late-time spectra and/or a comparison between the two. We find that asymmetric, possibly disk-like CSM, quickly overrun by ejecta, with filling factors on the order of a few per cent, similar to SN~2021irp \citep{reynolds25}, is compatible with a large fraction of the sample. Disk-like CSM configurations would be a natural consequence of mass loss through binary interaction.
    \item Out of the SNe with late-time spectra available, only in SNe~2019zcr and 2019uba does a spherical CSM remain plausible. Both of these are energetic enough in terms of total radiated energy to seemingly require a central engine even so. However, SN~2008es, which also radiated $\gtrsim10^{51}$~erg \citep{miller09}, shows signs of a filling factor $f\ll1$, further strengthening the need for an engine. 
    \item At the same time, this also showcases the diversity of the SLSN~II class, even within the most luminous group connected by the need for a central engine, some of which exhibit a more extended CSM and longer-lasting interaction than others. It is possible that the diversity within the class is connected to different viewing angles, and/or binary configurations that result in different mass-loss histories.
    \item SLSN~II progenitors show high mass-loss rates despite the lack of narrow lines. The UV excess seen in their SEDs, if assumed to be the sole outlet of CSI power, requires rates between $\sim10^{-4}\, \mathrm{M_\odot}$~yr$^{-1}$ and $\sim0.03\, \mathrm{M_\odot}$~yr$^{-1}$ in the years and decades before the SN. 
    \item These numbers should be treated as conservative lower limits: the excess luminosity co-evolves with the optical luminosity, implying that a large fraction of the interaction luminosity in fact emerges in the optical. This co-evolution also implies that the source of the UV excess is linked to the source of the optical luminosity and not a "secondary" interaction with a tenuous outer CSM.
\end{itemize}

For future SLSN~II observations, we stress the importance of late-time data, multi-wavelength coverage and polarimetry not only for the purpose of studying the mass-loss history, but also for determining the presence of a central engine. In the era of the Legacy Survey of Space and Time, large numbers of faint, high-redshift SLSNe~II will become available to the community, but sample size is not everything. Detailed studies of nearby events remain crucial for understanding their properties and for interpreting the populations of distant SLSNe. 

\begin{acknowledgements}

We thank Peter Lundqvist, Jesper Sollerman, Keiichi Maeda and Claes Fransson for helpful discussions.

T.K. and U.P. acknowledge support from the Research Council of Finland project 360274. \\
      
\end{acknowledgements}

%

\bibliographystyle{aa}
\bibliography{kangas_slsnII_20261012}
\clearpage

\begin{appendix}




\section{Supplementary tables}
\onecolumn

\FloatBarrier

\begin{table*}
\centering
\caption{Properties of the sample and available data.}
\label{tab:coverage}
\begin{tabular}{ccccccc}
\hline
SN & $z$ & Rest-frame $g$ peak & NUV coverage & Optical filters & NIR coverage & Late-time spectra \\
  &  & (MJD) & (d) & (observer-frame) & (observer-frame, d) & (d) \\
\hline
2008es & 0.205 & 54600.2 & +0 to +54 & $UBgVrRiI$ & +1 to +26, +254 to +302 & - \\
2018jkq & 0.119 & 58475.5 & - & $gri$ & - & +170 \\
2018lqi & 0.202 & 58783.7 & - & $gr$ & - & - \\
2019cqc & 0.117 & 58595.2 & +27 to +53 & $UBgVriz$ & - & +412, +488 \\
2019gsp & 0.171 & 58636.1 & - & $gri$ & - & - \\
2019kwr & 0.202 & 58589.0 & - & $griz$ & - & - \\
2019pud & 0.114 & 58754.6 & +14 to +18 & $UBgVriz$ & - & +240, +293, +341 \\
2019uba & 0.304 & 58810.2 & +10 to +16 & $UBgVriz$ & - & +209, +343 \\
2019xfs & 0.116 & 58809.6 & +33 to +73 & $UBgVr$ & - & +189, +297 \\
2019zcr & 0.260 & 58901.4 & $-13$ to +48 & $UuBgVriz$ & - & +264 \\
2019aanx & 0.403 & 58828.4 & - & $gr$ & - & - \\
2020bfe & 0.099 & 58919.5 & - & $griz$ & - & +171, +195, +306 \\
2020hgr & 0.126 & 58990.7 & $-5$ to +39 & $UuBgVriz$ & - & +236 \\
2020jhm & 0.057 & 58993.1 & +22 to +36 & $UBgVri$ & - & - \\
2023gpw & 0.083 & 60085.0 & $-8$ to +52 & $ugriz$ & +23 & - \\
\hline
\end{tabular}
\tablefoot{Dates are relative to the rest-frame $g$-band peak of each target. The peak epoch of SN~2008es is based on our GP interpolation (see Sect.~\ref{sec:BBfits}), while the rest are from \citetalias{kangas22} or, in the case of SN~2023gpw, \citet{kangas26}. Note that the epochs listed in \citet{hebbar26} for SN~2019cqc differ from these, as they are based on the $r$ rather than $g$ band.}
\end{table*}

\begin{table}
\centering
\caption{\textit{WISE} IR photometry of our sample.}
\label{tab:wisemags}
\begin{tabular}{ccccccc}
\hline
SN & MJD & Epoch & $W1$ & Error & $W2$ & Error \\
  &  & (d) & (mag, Vega) & (mag, Vega) & (mag, Vega) & (mag, Vega) \\
\hline
2018jkq & 58319.1 & $-139.8$ & $>17.85$ & - & $>16.79$ & - \\
2018jkq & 58479.6 & 3.7 & $>17.82$ & - & $>16.62$ & - \\
2018jkq & 58685.4 & 187.6 & $>18.12$ & - & $>16.62$ & - \\
2018jkq & 58847.7 & 332.5 & $>17.99$ & - & $>16.71$ & - \\
2018jkq & 59050.5 & 513.7 & $>17.77$ & - & $>16.87$ & - \\
\hline
2018lqi & 58678.6 & $-87.4$ & $>18.36$ & - & $>16.77$ & - \\
2018lqi & 58839.4 & 46.3 & $>17.92$ & - & $>16.87$ & - \\
2018lqi & 59044.5 & 217.0 & $>18.17$ & - & $>16.84$ & - \\
2018lqi & 59205.0 & 350.5 & $>18.30$ & - & $>16.95$ & - \\
2018lqi & 59410.0 & 521.1 & $>17.89$ & - & $>16.54$ & - \\
\hline
2019cqc & 58570.3 & $-22.3$ & 17.87 & 0.10 & $>17.05$ & - \\
2019cqc & 58734.2 & 124.5 & 17.25 & 0.09 & $>17.27$ & - \\
2019cqc & 58936.0 & 305.1 & 17.98 & 0.14 & $>16.69$ & - \\
2019cqc & 59100.0 & 451.9 & 17.15 & 0.10 & $>16.79$ & - \\
2019cqc & 58303.3 & 633.9 & 18.02 & 0.17 & $>16.94$ & - \\
\hline
2019gsp & 58436.7 & $-170.3$ & $>18.17$ & - & $>16.69$ & - \\
2019gsp & 58639.4 & 2.8 & $>17.78$ & - & $>16.40$ & - \\
2019gsp & 58802.3 & 142.0 & $>17.97$ & - & $>16.70$ & - \\
2019gsp & 59003.5 & 313.8 & $>18.19$ & - & $>16.43$ & - \\
2019gsp & 59166.5 & 452.9 & $>18.22$ & - & $>16.73$ & - \\
2019gsp & 59369.2 & 626.0 & $>17.84$ & - & $>16.49$ & - \\
\hline
2019kwr & 58464.7 & $-103.4$ & $>18.25$ & - & $>17.25$ & - \\
2019kwr & 58623.5 & 28.7 & $>18.22$ & - & $>16.89$ & - \\
2019kwr & 58831.5 & 201.7 & $>18.41$ & - & $>17.22$ & - \\
2019kwr & 58987.8 & 331.7 & $>18.35$ & - & $>17.07$ & - \\
2019kwr & 59196.0 & 504.9 & $>18.27$ & - & $>17.19$ & - \\
\hline
2019pud & 58609.6 & $-130.1$ & $>17.86$ & - & $>16.77$ & - \\
2019pud & 58773.5 & 17.0 & 17.33 & 0.12 & $>16.53$ & - \\
2019pud & 58973.9 & 196.8 & $>17.77$ & - & $>16.39$ & - \\
2019pud & 59137.7 & 343.9 & $>17.97$ & - & $>16.48$ & - \\
2019pud & 59339.4 & 525.0 & $>17.86$ & - & $>16.38$ & - \\
\hline
\end{tabular}
\tablefoot{Epochs are relative to the rest-frame $g$-band peak of each target.}
\end{table}

\setcounter{table}{1}

\begin{table}
\centering
\caption{\textit{WISE} IR photometry of our sample (continued).}
\label{tab:wisemags}
\begin{tabular}{ccccccc}
\hline
SN & MJD & Epoch & $W1$ & Error & $W2$ & Error \\
  &  & (d) & (mag, Vega) & (mag, Vega) & (mag, Vega) & (mag, Vega) \\
\hline
2019uba & 58690.8 & $-91.6$ & $>17.93$ & - & $>16.74$ & - \\
2019uba & 58850.9 & 31.2 & $>17.86$ & - & $>16.74$ & - \\
2019uba & 59056.5 & 188.9 & $>17.92$ & - & $>16.74$ & - \\
2019uba & 59216.6 & 311.7 & $>18.07$ & - & $>16.74$ & - \\
2019uba & 59422.9 & 470.0 & $>18.24$ & - & $>16.74$ & - \\
\hline
2019xfs & 58589.0 & $-197.7$ & $>17.91$ & - & $>16.95$ & - \\
2019xfs & 58757.2 & $-46.9$ & 17.73 & 0.14 & $>16.60$ & - \\
2019xfs & 58954.6 & 129.9 & $>18.36$ & - & $>17.20$ & - \\
2019xfs & 59118.7 & 276.9 & $>18.01$ & - & $>17.00$ & - \\
2019xfs & 59320.3 & 457.6 & $>17.90$ & - & $>17.09$ & - \\
2019xfs & 59485.0 & 605.2 & $>17.82$ & - & $>16.88$ & - \\
\hline
2019zcr & 58846.5 & $-43.6$ & $>18.09$ & - & $>16.98$ & - \\
2019zcr & 59006.3 & 83.3 & $>17.82$ & - & $>16.86$ & - \\
2019zcr & 59210.8 & 245.5 & $>18.08$ & - & $>17.14$ & - \\
2019zcr & 59370.4 & 372.3 & $>18.41$ & - & $>16.82$ & - \\
2019zcr & 59576.5 & 535.8 & $>17.93$ & - & $>17.23$ & - \\
\hline
2019aanx & 58773.3 & $-39.3$ & $>17.87$ & - & $>16.50$ & - \\
2019aanx & 58927.7 & 70.7 & $>17.93$ & - & $>16.35$ & - \\
2019aanx & 59137.5 & 220.3 & $>17.76$ & - & $>16.38$ & - \\
2019aanx & 59293.6 & 331.6 & $>18.00$ & - & $>16.80$ & - \\
2019aanx & 59501.6 & 479.8 & $>17.86$ & - & $>16.47$ & - \\
2019aanx & 59658.9 & 592.0 & $>17.82$ & - & $>17.00$ & - \\
\hline
2020bfe & 58729.1 & $-173.2$ & $>18.06$ & - & $>17.02$ & - \\
2020bfe & 58929.6 & 9.2 & 16.90 & 0.08 & - & - \\
2020bfe & 59091.9 & 156.8 & 17.46 & 0.11 & $>17.01$ & - \\
2020bfe & 59293.8 & 340.6 & $>18.03$ & - & - & - \\
2020bfe & 59457.6 & 489.6 & $>18.03$ & - & $>17.13$ & - \\
\hline
2020hgr & 58805.9 & $-164.2$ & $>19.13$ & - & $>17.39$ & - \\
2020hgr & 58965.0 & $-22.9$ & $>18.60$ & - & $>17.27$ & - \\
2020hgr & 59171.5 & 160.6 & $>18.80$ & - & $>17.10$ & - \\
2020hgr & 59327.9 & 299.5 & $>19.12$ & - & $>17.90$ & - \\
2020hgr & 59538.8 & 486.8 & $>18.84$ & - & $>17.33$ & - \\
\hline
2020jhm & 58817.1 & $-166.5$ & $>18.84$ & - & $>17.03$ & - \\
2020jhm & 58974.1 & $-18.0$ & 18.00 & 0.14 & $>17.11$ & - \\
2020jhm & 59183.8 & 180.4 & 17.22 & 0.10 & 15.85 & 0.05 \\
2020jhm & 59339.5 & 327.7 & 17.45 & 0.11 & 16.82 & 0.09 \\
2020jhm & 59546.9 & 524.0 & $>18.92$ & - & 17.22 & 0.10 \\
2020jhm & 59700.9 & 669.6 & $>18.72$ & - & $>17.39$ & - \\
\hline
2023gpw & 59954.4 & $-120.6$ & $>17.91$ & - & $>17.11$ & - \\
2023gpw & 60111.3 & 24.3 & 16.48 & 0.07 & $>16.60$ & - \\
2023gpw & 60318.2 & 215.4 & $>18.16$ & - & $>16.72$ & - \\
2023gpw & 60477.1 & 362.1 & $>18.05$ & - & $>16.67$ & - \\
\hline
\end{tabular}
\tablefoot{Epochs are relative to the rest-frame $g$-band peak of each target.}
\end{table}

\FloatBarrier

\section{Blackbody fit plots}

\begin{figure*}
\centering
\includegraphics[width=0.328\linewidth]{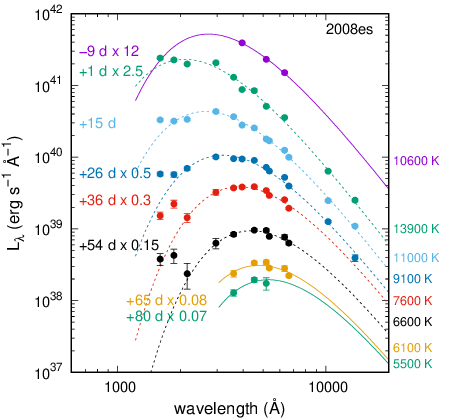}
\includegraphics[width=0.328\linewidth]{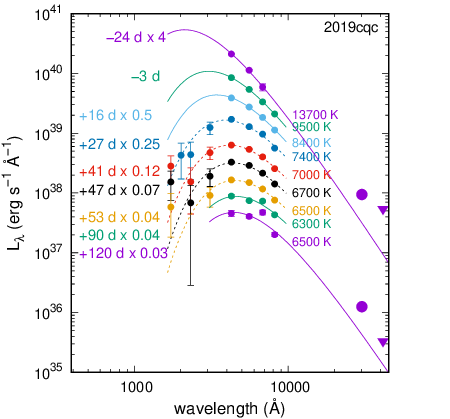}
\includegraphics[width=0.328\linewidth]{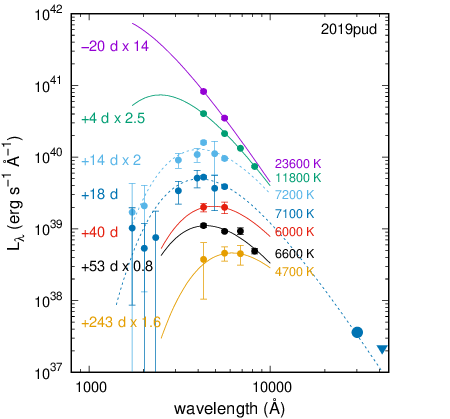}
\includegraphics[width=0.328\linewidth]{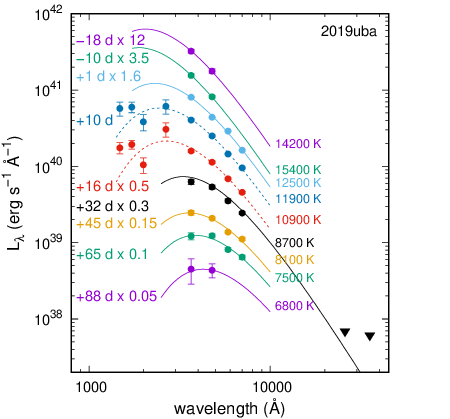}
\includegraphics[width=0.328\linewidth]{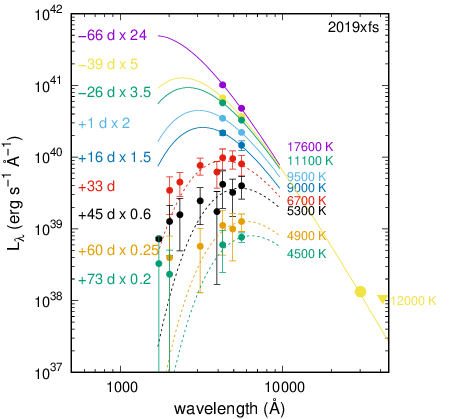}
\includegraphics[width=0.328\linewidth]{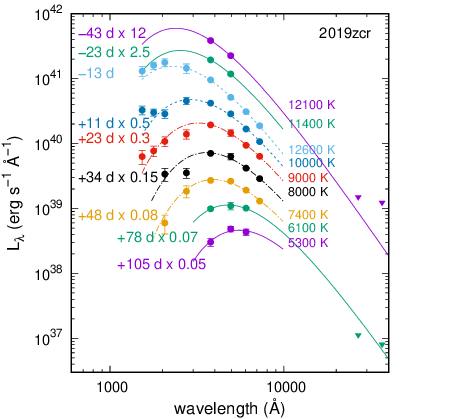}
\includegraphics[width=0.328\linewidth]{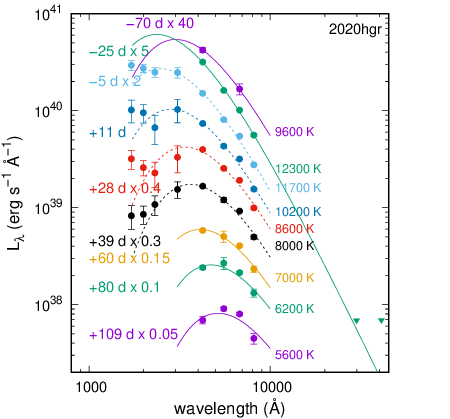}
\includegraphics[width=0.328\linewidth]{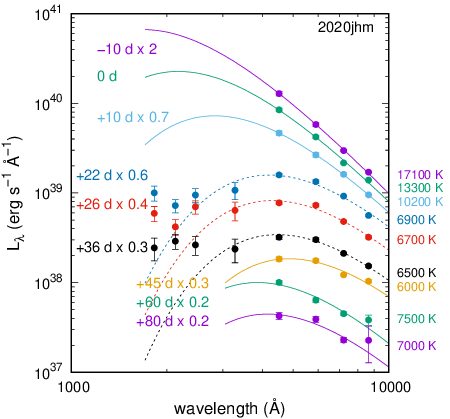}
\includegraphics[width=0.328\linewidth]{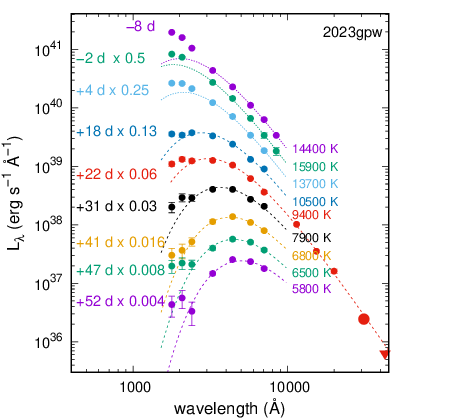}
\caption{Blackbody fits (lines) to the UV, optical and/or IR SEDs (points) of those events in our sample with more than three bands available for the fitting process. Solid lines indicate all plotted points except $W1$ and $W2$ were used; dashed lines indicate $UVW2$ and $UVM2$ were also ignored; and dotted lines indicate $UVW1$ was ignored as well.}
\label{fig:BB}
\end{figure*}

\begin{figure*}
\centering
\includegraphics[width=0.4\linewidth]{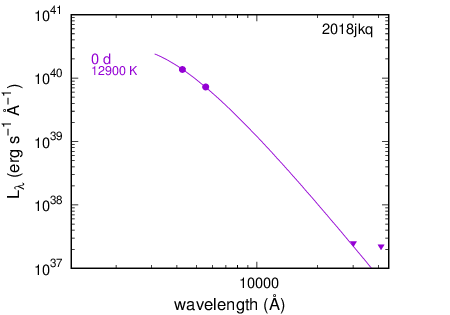}
\includegraphics[width=0.4\linewidth]{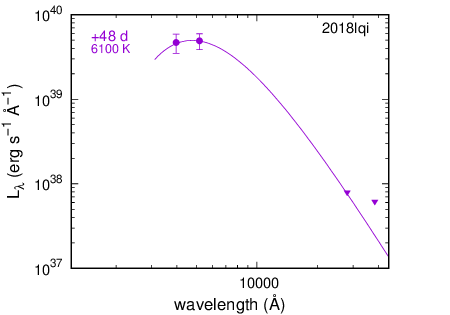}
\includegraphics[width=0.4\linewidth]{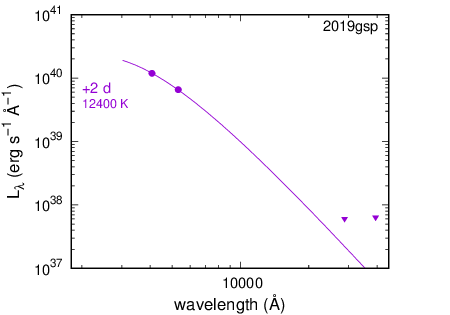}
\includegraphics[width=0.4\linewidth]{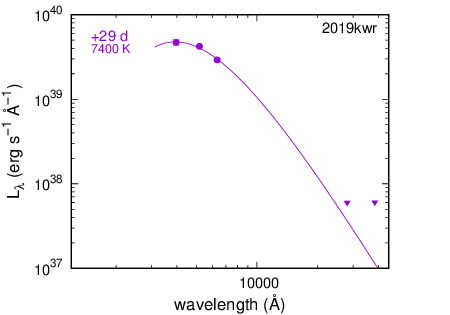}
\includegraphics[width=0.4\linewidth]{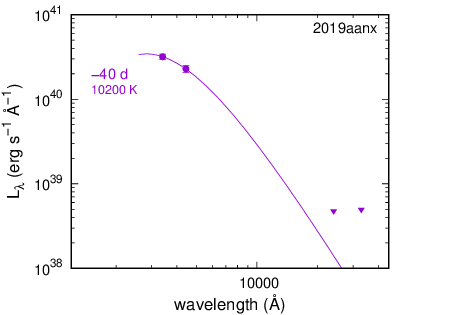}
\includegraphics[width=0.4\linewidth]{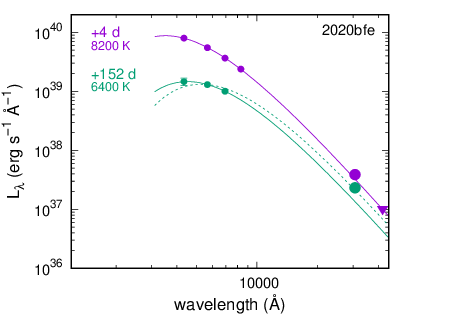}
\caption{Blackbody fits (lines) to the UV, optical and/or IR SEDs (points) of those events in our sample with only two or three bands available for the fitting process in most or all epochs. Solid lines indicate $W1$ and $W2$ were not used in the fit; the dashed line indicates $W1$ was used in order to test the significance of the possible late-time IR excess in SN~2020bfe.}
\label{fig:BB2}
\end{figure*}

\end{appendix}
\end{document}